# Trapped modes with extremely high quality factor in the subwavelength ring resonator composed of dielectric nanorods

Hai-bin Lü[1], Xiaoping Liu[2,*]

[1]*College of Optoelectronic Science and Engineering, National University of Defense Technology, Changsha, Hunan 410073, China*
[2] *ShanghaiTech University, Shanghai, Shanghai 201210, China*

**Abstract:** In this paper, we investigate numerically the trapped modes with near zero group velocities supported in the ring array composed of dielectric nanorods, based on a two-dimensional model. Two sorts of trapped modes in the ring array have been identified: the BCR trapped modes which correspond to the bound modes below the light line at the edge of the first Brillouin zone in the corresponding planar structure (namely the infinite linear chain); the quasi-BIC trapped modes corresponding to the bound states in the continuum supported in the infinite linear chain. According to the whispering gallery condition, the BCR trapped modes can be supported in the ring array only when the number of dielectric elements $N$ is even, while the quasi-BIC ones always exist no matter whether $N$ is odd or even. For both two kind of trapped modes, the lowest one of each kind possesses the highest Q factor, which are $\sim 10^5$ for BCR kind and $\sim 10^{11}$ for quasi-BIC kind with $N=16$ respectively, and the radiation loss increases dramatically as the structural resonance increases. Finally, the behavior of the Q factor with $N$ is explained numerically for the lowest one of each kind of trapped modes. The Q factor scales as $Q \sim \exp(0.662N)$ for the quasi-BIC trapped mode and $Q \sim \exp(0.325N)$ for the BCR one. Intriguingly, the Q factor of the quasi-BIC trapped mode can be as large as $\sim 10^5$ even at $N=8$. Compared to the finite linear chain, the structure of ring array exhibits overwhelming advantage in Q factor with the same $N$ because there is no array-edge radiation loss in the ring array. We note that the principles can certainly be extended to particles of other shapes (such as nanospheres, nanodisks, and many other experimentally feasible geometries).

# Introduction

When light is brought to a standstill, its interaction with media increases dramatically due to a singularity in the density of optical states. Meanwhile, small group velocity increases the round-trip travel time inside the cavity and hence results in a large quality factor Q. Thus, realizing bound modes with near zero group velocity in resonators has meaningful potentials in practical applications, such as low-threshold nanolasers, enhanced light-matter interactions and strong-coupling cavity quantum electrodynamics [1-8]. In order to obtain small group velocities, a slow light regime needs to be introduced, which could be achieved effectively, for example, in photonic crystal waveguides via coherent backscattering [9-10], through symmetry breaking in axially uniform waveguides [11-12], in metamaterials through the plasmonic analogue of electromagnetically induced transparency [13], or in planar plasmonic heterostructures [14-15]. Utilizing the aforementioned slow-light mechanisms, researchers have realized resonant modes with small or even near zero group velocities in resonators based on the axially uniform waveguides, planar plasmonic heterostructures or photonic crystal waveguides and microcavities [12,16-18].

Moreover, one-dimensional chains of high-index dielectric nanoparticles have been intensively studied in recent years because of their abilities of guiding and confining light in the nanoscale and tailoring the dispersion relations to achieve slow or trapped light in analogy to photonic crystals due to the discrete translational symmetry [19-26]. Compared to plasmonics, this system has significantly lower material losses and more degree of freedom (such as electric and magnetic dipole resonances) to engineer the optical features of systems. Compared to photonic crystal waveguides, fewer materials are used in dielectric chains to achieve light guiding and dispersion engineering in the nanoscale. Moreover, remarkable field enhancement and confinement are also observed in the dielectric nanoparticle chains because the individual elements are resonant at the wavelength of interest [26]. Generally speaking，light confinement and guiding in the 1D infinite linear nanoparticle chain are obtained by exciting the bound modes which are below the light line of the host medium. Hence, light guiding along the chain is guaranteed by the totally internal reflection [22]. On the other hand，for 1D infinite linear chains，there also exist bound states in the continuum (BICs) and corresponding quasi-BIC propagation bands above the light line [27-29]. These BIC modes correspond to exceptional points in the quasi-BICs propagation bands where the imaginary part of the resonant frequency turns to zero. In other words, these BIC modes possess an infinite Q factor. However, in practice, one always deals with finite arrays. For the finite arrays, one can expect that the highest quality factor can be attained for bound modes in an array of particles arranged in a circle and equidistant from each other because a circle has no sharp ends and resulting edge radiations in contrast with a particle chain [20]. On the base of the photonic bands below and above the light line in the infinite linear chains of dielectric nanoparticles，here we propose to employ the ring array of high dielectric nanorods to realize resonator which is expected to support near zero-group-velocity trapped modes with high Q factors.

## Results and Discussions

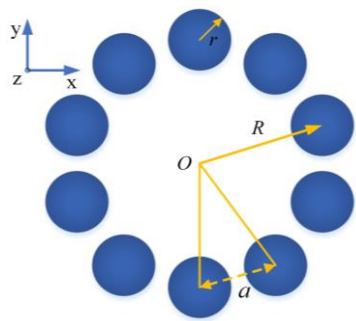

Fig 1 Schematic diagram of the ring array of dielectric nanorods.

To describe the electromagnetic resonances of multi-particle systems, a valid approach is the multiple-particle Mie scattering theory or the simplified coupled dipole model [20,21,24,30]. In the multiple-scattering method, the scattered field is expanded over a series of multipolar response functions weighted with scattering amplitudes, and consequently the problem of electromagnetic resonances of a multi-particle system can be converted to the eigenvalue problem about the scattering amplitudes without the incident field considered. Hence, one can obtain the information about the eigenmodes by solving the dispersion equation. Using the aforementioned method, Ref. 20 investigated the eigenmodes supported in circular arrays of dielectric nanospheres. However, because of the dipole approximation used in the numerical procedure to simplify the cumbersome computations, the calculated results are only well justified for eigenmodes corresponding to dipolar Mie resonances possessing the lowest frequency, but cannot describe any quasi-BIC eigenmodes corresponding to the multipolar Mie resonances [28]. Here, for obtaining reliable results for bound modes with higher eigenfrequencies, we employ the finite-element method to perform full-wave calculation for the problem of structural resonances of the ring array. In this paper, a ring array composed of *N* 2D circular elements is considered. Such 2D circular elements serve as models of 3D nanorods which uniformly extend in the *z* direction. According to Ref. 20, high quality modes in an array of ten or more particles can be attained at least for a refractive index n>2. Thus, the parameters used in the numerical calculation are set as follows: the particle number *N*=16，the radius of each element *r*=120nm, the distance between the adjacent elements *a*=280nm, and the radius of the ring array *R*=a/2sin(π/*N*). The nanorods with refractive index n=3.5 are embedded in air. The material loss and dispersion are neglected in the calculation.

The property of the ring array shown in Fig.1 can be understood well by first analyzing a corresponding planar structure, that is, an infinite linear nanoparticle chain with the same geometry parameters *r* and *a*. With the modes in the corresponding linear chain, the resonances of the ring array can be understood using the whispering gallery condition [31]. The ring array supports the *m*th order resonance at a frequency *ω* when the Bloch vector *β* of a mode in the corresponding planar structure satisfies:

$$\beta \cdot 2\pi R = 2\pi m \qquad (1)$$

Where *m*=0,1,2,…represents the angular momentum of an eigenmode in the ring array. Because the infinite linear chain possesses bound and quasi-bound propagation bands below and above the light line, the ring array should inherently support two sorts of bound eigenmodes which are located below and above the light line respectively. The detailed calculations presented later show that there indeed exist the aforementioned two kind of bound modes. The ones below the light line are referred as below-continuum-resonance (BCR) bound modes, and the ones above the light line are quasi-BIC bound modes whose traces could be observed in form of high-Q structural resonances although formally BICs do not exist in finite arrays [29]. Next, we will investigate these two kind of bound modes in detail, and suppose that only the transverse electric (TE, with the electric field polarized along the $z$ axis) electromagnetic eigenmodes are considered.

**BCR trapped modes**

First, the bound modes below the light line are considered. We plot the corresponding resonant frequencies and the leakage rates of the ring array as a function of angular momentum $m$ along with the below-the-light-line TE photonic bands of the infinite linear chain in Fig. 2(a).

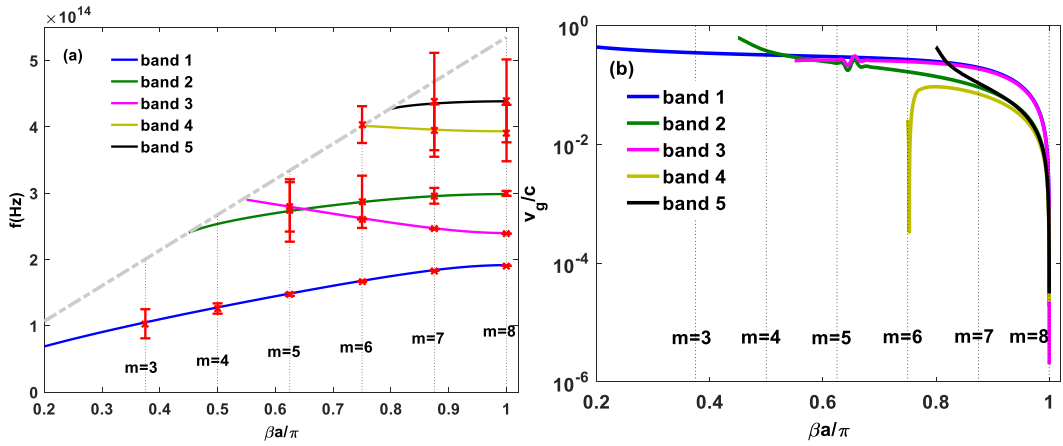

Fig. 2 (a) The resonant frequencies and the leakage rates of the ring array as a function of angular momentum $m$ along with the TE photonic bands of the infinite linear chain. For the infinite linear chain, the radius of each circular element and the period are 120nm and 280nm, respectively. The grey dash-dot line is the light line of air. The bars represent the properties of the structural resonances in the ring array at different angular momentums labeled by $m$. The center (marked by red crosses) and the height of each bar correspond to the resonant frequency and the leakage rate, respectively. (b) The group velocity distributions for different photonic bands in (a).

As shown in Fig. 2(a), it is clear that all the resonant frequencies of the ring array distribute along the corresponding photonic bands of the infinite linear chain, and the positions of these structural resonances, agree very well with Eq. (1) for different angular momentums $m$ when $N$=16. Moreover, the leakage rates of these structural resonances decease dramatically along the photonic bands with increasing the angular momentum, which results from continually going away from the light line. On the other hand, at a certain angular momentum *m*, the larger the resonant frequency of the ring array is, the larger the corresponding leakage rate is, because it's more close to the light line and hence the radiation loss increases remarkably.

Fig. 2(b) presents the distributions of group velocity for different TE photonic bands. Obviously, due to the symmetries of photonic bands of systems having discrete translational symmetry, the group velocity goes to zero as the Bloch wavevector *β* approaches π/a, the band edge in the first Brillouin zone. On the other hand, when *β*=π/a, the corresponding angular momentum of the guided modes in the ring array *m*=*βR*≈*N*/2 according to Eq.(1). Hence, *m* is integer around the band edge only when *N* is even, which is verified in Fig. 2(a) where *N*=16 and *m*=8. Thus, one can expect that the ring array with even *N* supports structural resonances with near zero group velocities in the vicinity of the band edge below the light line. In Table 1 we collect the parameters of the five band-edge BCRs with *m*=8 shown in Fig. 2(a). Among these parameters, the group velocity of a certain mode is calculated using the ratio of the energy flux to the energy density [32]:

$$\vec{v}_g = \frac{\int \vec{S} d^3 \vec{r}}{\int w d^3 \vec{r}} \quad (2)$$

Where $\vec{S}$ and *w* are the averaged energy flux and energy density of the mode. The integrations are performed over the entire ring array.

Table 1. Band-edge BCRs in the ring array of *N* 2D circular elements. N=16, r=120nm, a=280nm, n=3.5. The Q factor is calculated by using the standard definition Q=ω/Γ, where ω and Γ are the resonant angular frequency and leakage rate respectively. $\bar{v}_g^\varphi$ is the averaged azimuthal group velocity over the whole ring array.

| Band-edge BCR | $f_{eigen}$ (Hz) | Q factor | $\bar{v}_g^\varphi$ (m/s) |
|---|---|---|---|
| 1 | $1.91 \times 10^{14}$ | $1.42 \times 10^5$ | $1.45 \times 10^{-4}$ |
| 2 | $2.39 \times 10^{14}$ | $8.58 \times 10^3$ | $1.25 \times 10^{-2}$ |
| 3 | $2.99 \times 10^{14}$ | 517.4 | 0.97 |
| 4 | $3.91 \times 10^{14}$ | 57.7 | 14.8 |
| 5 | $4.39 \times 10^{14}$ | 44.2 | 24.9 |

As is shown in Table 1, all the five band-edge BCRs have near zero group velocities, proving that the ring array with even *N* can support eigenmodes with near zero group velocity at the band edge. Also, the group velocity increases dramatically when the mode approaches to the light line, and the lowest mode can obtain the group velocity down to $10^{-4}$ m/s. Hence, these near zero-group-velocity modes are referred as BCR trapped modes. Moreover, it is also shown that the BCR trapped mode experiences more and more radiation loss when it gets close to the light line while the one possessing the lowest frequency presents the highest Q factor up to $10^5$. Therefore, one can conclude that the lowest BCR trapped mode has not only the highest Q factor but also the smallest group velocity. Fig. 3 presents the normalized electromagnetic field patterns for the lowest BCR trapped mode. It can be observed that the trapped mode is in antiphase, which means that the neighboring elements can cancel the multipolar radiation fields from each other so as to obtain significant suppression of radiation loss [30].

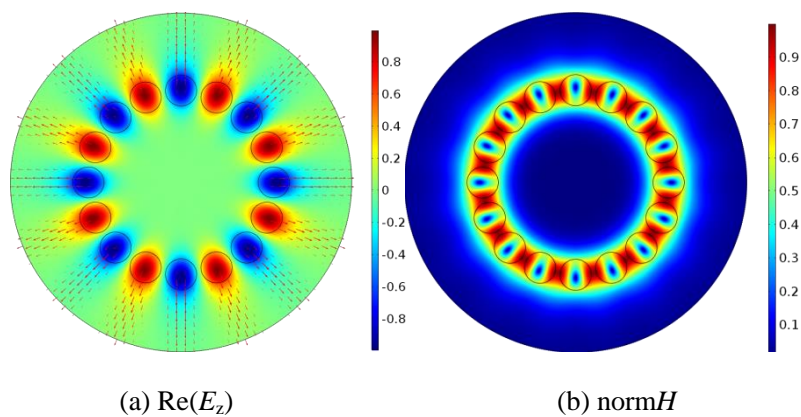

(a) Re($E_z$)  (b) norm*H*

Fig. 3. Electromagnetic field patterns for the lowest BCR trapped mode. (a) real part of $E_z$; (b) norm*H*. Red arrows in the $E_z$ pattern show the Poynting vectors carrying the power flow towards the outside, which means the radiation loss.

**Quasi-BIC trapped modes**

Next, the quasi-BIC bound modes are considered. Similarly, the corresponding resonant frequencies and the leakage rates of the ring array are presented in Fig. 4 along with the TE quasi-BIC propagation bands of the infinite linear chain.

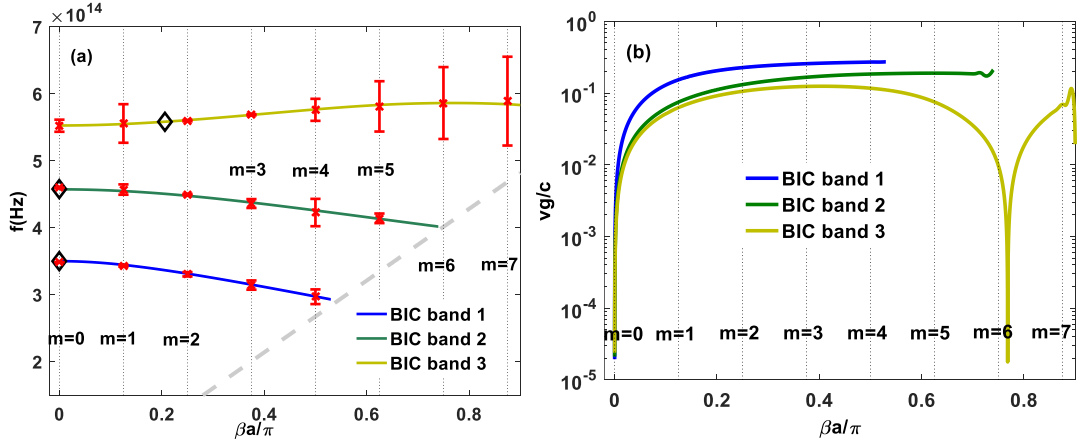

Fig.4. (a) The corresponding resonant frequencies and the leakage rates of the ring array as a function of angular momentum m along with the three lowest TE quasi-BIC propagation bands of the infinite linear chain. For the infinite linear chain, the radius of each circular element and the period are 120nm and 280nm, respectively. The grey dash line is the light line of air. The Bloch BICs supported by the infinite linear chain are marked by black diamonds. The red bars represent the properties of the structural resonances in the ring array at different angular momentums m. The center (marked by crosses) and the height of each bar correspond to the resonant frequency and the leakage rate, respectively. (b) The group velocity distributions for different quasi-BIC propagation bands in (a).

Fig. 4(a) presents the three lowest TE quasi-BIC propagation bands above the light line in the infinite linear chain, corresponding to two Bloch BICs with $\beta=0$ and one Bloch BIC with $\beta\neq 0$ which are all marked by black diamonds, and shows clearly that positions of the structural resonances of the ring array agree very well with Eq. (1) for different angular momentums m along the corresponding quasi-BIC bands when $N=16$. Although there are no exact BICs in finite arrays, their traces could be observed in form of high-Q structural resonances in the ring array due to the discrete rotational symmetry. For the two Bloch BICs with $\beta=0$, the corresponding angular momentum of the guided modes in the ring array $m=\beta R=0$, and hence there exist two corresponding structural resonances of the ring array at the almost exact positions of the two BICs (marked by black diamonds at $\beta=0$), which is shown clearly in Fig. 4(a). On the other hand, for the Bloch BIC with $\beta\neq 0$, the corresponding angular momentum of the guided modes in the ring array $m=\beta R\approx 1.65$, which violates the whispering gallery condition, and hence no corresponding structural resonance exists in its position as is shown in Fig. 4(a). However, because $R=a/2\sin(\pi/N)$, it can be expected that there can exist the corresponding structural resonance for the BIC with

$\beta \neq 0$ as long as one adjusts the *N* to make sure the angular momentum *m* is a integer. Moreover, besides these aforementioned quasi-BIC structural resonances, there are also other structural resonances of different *m* supported in the ring array, departing from these three Bloch BIC points along the photonic bands. However, it should be pointed out that, as a result of dramatically increased coupling between the open diffraction channels and the bound modes [27], the leakage rates of structural resonances would increase abruptly once they depart from the corresponding BIC points, making them less meaningful in practical application than those quasi-BIC structural resonances.

Fig. 4(b) presents the group velocities for different TE quasi-BIC propagation bands. Because of the time-reversal symmetry of photonic bands of Hermitian systems, the group velocity always goes to zero as the Bloch wavevector $\beta$ approaches 0. On the other hand, when $\beta=0$, the corresponding angular momentum of the guided modes in the ring array $m=0$. Thus, one can expect that the ring array with arbitrary *N* always supports quasi-BIC bound modes with near zero group velocity at $m=0$. In Table 2 we collect the parameters of the two quasi-BIC bound modes with $m=0$ shown in Fig. 4(a).

Table 2. Quasi-BIC bound modes with $m=0$ in the ring array of *N* 2D circular elements. $N=16$, $r=120$nm, a=280nm, n=3.5. $\bar{v}_g^{\varphi}$ is the averaged azimuthal group velocity over the whole ring array.

| Quasi-BIC Mode | $f_{eigen}$ (Hz) | Q factor | $\bar{v}_g^{\varphi}$ (m/s) |
|---|---|---|---|
| 1 | $3.49 \times 10^{14}$ | $1.30 \times 10^{11}$ | $3.34 \times 10^{-2}$ |
| 2 | $4.60 \times 10^{14}$ | $3.85 \times 10^{7}$ | $1.99 \times 10^{-1}$ |

As is shown in Table 2, both the two quasi-BIC bound modes have near zero group velocities, and can be referred as trapped modes. Also, the group velocity increases evidently when the resonant frequency increases. Moreover, it is also shown that the trapped mode experiences more radiation loss when it gets higher resonant frequency, which results from the increased coupling between the open diffraction channels and the quasi-BIC bound modes, and the trapped mode possessing the lowest frequency presents the highest Q factor up to $10^{11}$. Therefore, one can conclude that the lowest quasi-BIC trapped mode has the highest Q factor among all the bound modes supported in the ring array. Fig. 5 presents the normalized electromagnetic field patterns for the lowest quasi-BIC trapped mode. It is shown that the Poynting vectors form energy vortices inside and outside each element, resulting in the radiation loss suppressed remarkably.

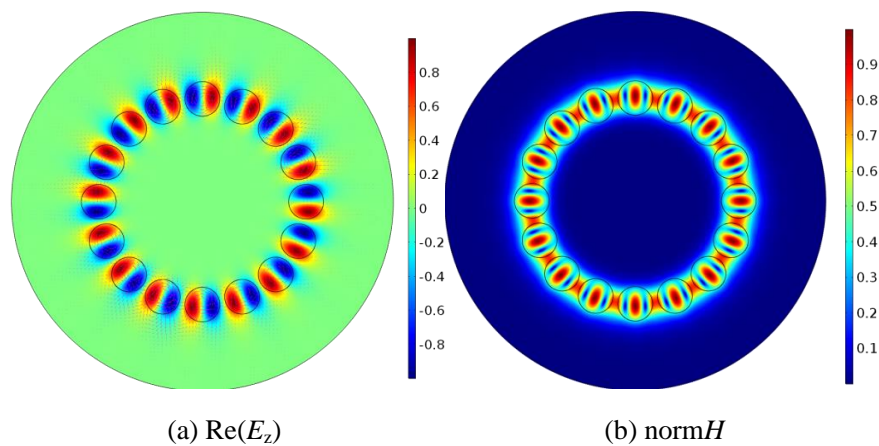

(a) Re($E_z$)  (b) norm$H$

Fig.5. Electromagnetic field patterns for the lowest quasi-BIC trapped mode. (a) real part of $E_z$; (b) norm$H$. Magenta arrows in the $E_z$ pattern show the Poynting vectors forming energy vortices inside and outside each element, which results to suppression of the radiation loss.

**Q factors against *N* of the ring array**

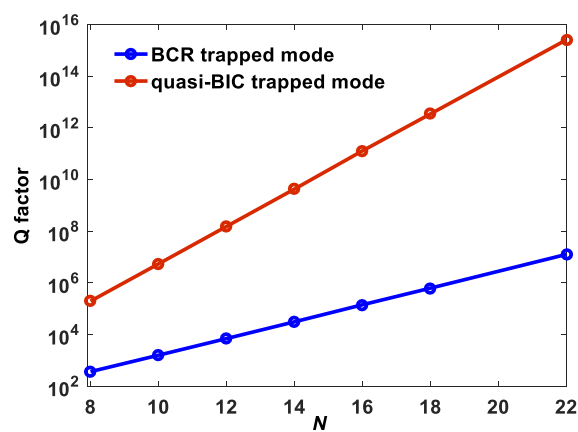

Fig.6. Q factors against the number of dielectric elements *N* in the ring array for the lowest one of each sort of trapped modes.

The Q factors against the number of dielectric elements in the ring array for the lowest one of each sort of trapped modes mentioned above are plotted in Fig. 6. As is depicted in Fig. 6, the Q factors for both two trapped modes experience exponential growth as *N* increases, and as linear fitted from the curves, the Q factor scales as Q~exp(0.662*N*) for the quasi-BIC trapped mode and Q~exp(0.325*N*) for the BCR trapped mode, indicating that the quasi-BIC trapped mode presents faster growth of Q factor than the BCR one. Also the Q factor of the quasi-BIC trapped mode is 3-8 orders of magnitude larger than that of the BCR one when *N*<20, and can be as large as ~$10^5$ even at *N*=8, which means that the quasi-BIC trapped mode is very attractive to realize low-threshold nanolasers. Moreover, compared to the finite linear array of dielectric elements [29], the structure of ring array exhibits the overwhelming advantage in Q factor with the same element number because there is no array-edge radiation loss in the ring array, which is a main radiation mechanism in finite linear arrays.

## Conclusion

In summary, we have demonstrated that the ring array composed of dielectric nanorods can support two kind of trapped modes with near zero group velocity and extremely high Q factor. The BCR ones correspond to the bound modes below the light line at the band edge in the corresponding infinite linear chain while the quasi-BIC ones correspond to the BICs supported above the light line. According to the whispering gallery condition, the BCR trapped modes can be supported in the ring array only when *N* is even, while the quasi-BIC tones always exist no matter whether *N* is odd or even. Moreover, the lowest quasi-BIC trapped mode has the highest Q factor, which can be as large as ~$10^5$ even at *N*=8, among all the bound modes supported in the ring array. Finally, the behavior of the Q factor with the number of elements in the ring array is explained numerically for the lowest one of each kind of trapped modes, and it suggests that the Q factors for both two trapped modes experience exponential growth as *N* increases, and the quasi-BIC trapped mode presents faster growth of Q factor than the BCR one. Because there is no array-edge radiation loss, the structure of ring array exhibits the overwhelming advantage in Q factor with the same *N* with respect to the finite linear chain. It should be noted that the principles obtained here can certainly be extended to particles of other shapes (such as nanospheres, nanodisks, and many other experimentally more feasible structures). Therefore, the ring-array structure of dielectric elements presents promising potentials in applications such as low-threshold nanolasers from BICs, tunable single-mode operations and enhanced light-matter interactions.